\documentstyle[preprint,aps,epsf]{revtex}
\begin{document}
\draft

\preprint{}
\title{Quasiparticle density of states in dirty high-$T_c$ superconductors}
\author{T. Senthil and Matthew P.A. Fisher}
\address{ Institute for Theoretical Physics, 
University of California,
Santa Barbara, CA 93106--4030  
}
 
\date{\today}
\maketitle
 
\begin{abstract}
We study the density of quasiparticle states of dirty $d$-wave superconductors.
We show the existence of singular corrections to the density of states due to quantum interference effects.
We then argue that the density of states actually vanishes in the localized phase as $|E|$ or $E^2$
depending on whether time reversal is a good symmetry or not. We verify this result for systems without 
time reversal symmetry
in one dimension using supersymmetry techniques. This simple, instructive calculation 
also provides the exact universal scaling function for the density of states for the
crossover from ballistic to localized behaviour in one dimension. Above two dimensions, we argue that
in contrast to the conventional Anderson localization transition, the density of states  has critical
singularities which we calculate in a $2+\epsilon$ expansion. We discuss consequences of our results for various 
experiments on dirty high-$T_c$ materials.
 
\end{abstract}
\vspace{0.15cm}


\narrowtext

\section{Introduction}
The question of the quasiparticle density of states of a 
two dimensional $d_{x^2-y^2}$
superconductor in the presence of disorder has been a matter of 
some controversy. Early theoretical work based on approximate
self-consistent treatments\cite{Lee,Gorkov,HPS} of the disorder demonstrate
that a finite Fermi level density of states is generated for
arbitrarily weak disorder. In contrast,  some exact results\cite{Tsvelik} for
a simplified model of the disorder which ignores
the scattering between the two pairs of antipodal nodal points
show that the density of states($\rho(E)$) vanishes on approaching
zero energy (measured from the Fermi energy) as 
$\rho(E) \sim E^{\frac{1}{7}}$. Claims of rigourous proofs\cite{Ziegler} of a 
constant non-zero density of states have also appeared 
in the literature. 

\begin{figure}
\epsfxsize=4.0in
\centerline{\epsffile{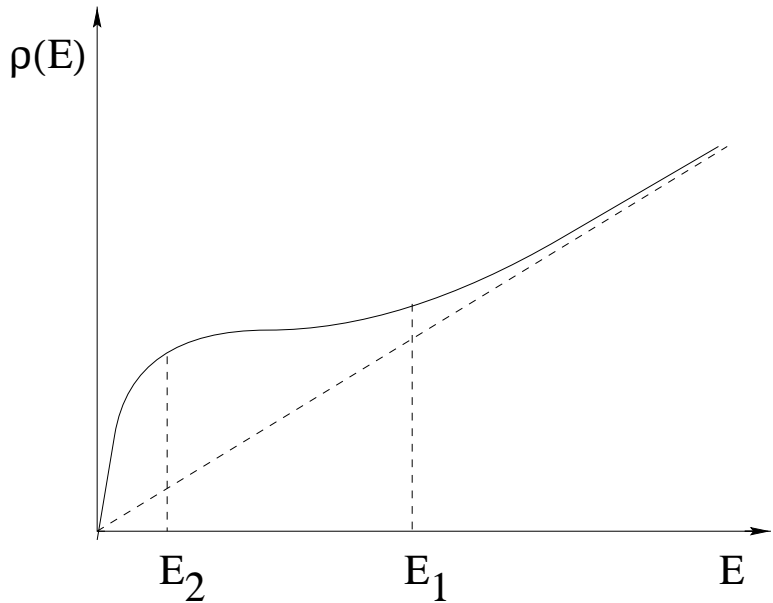}}
\vspace{0.05in}
\caption{ Density of states of the two dimensional dirty $d_{x^2 - y^2}$
superconductor
}
\vspace{0.04in}  
\label{dos2d}
\end{figure}

In a recent paper\cite{short}, we discussed the problem of quasiparticle transport 
and localization in dirty superconductors ignoring the quasiparticle interactions, and 
treating the disorder with a non-linear sigma model
field theory. The starting point for the sigma model description is 
the approximate self-consistent  treatment of the disorder which,
as mentioned above, generates a finite density of states. We
argued that inclusion of small harmonic fluctuations about the self-consistent
solution leads to diffusion of the spin and energy densities 
of the quasiparticles (though not of the charge density). 
Quantum interference effects finally lead to quasiparticle 
localization at the longest length scales in two dimensions.
In this paper, we consider the behaviour of the density of states
in the sigma model. We show that in the diffusive regime,
quantum interference effects lead to a singular logarithmic suppression
of the density of states. We then argue that in the localized spin 
insulator, the density of states actually 
vanishes as $|E|$ for superconductors with both spin rotation and
time reversal symmetry.  
A schematic plot of the density of states as a function of energy
is shown in Figure \ref{dos2d}. The linear density of states of the 
pure $d_{x^2-y^2}$ superconductor gets rounded off at an
energy scale $E_1$ of the order of the elastic scattering rate.
This marks the crossover from the ballistic to the diffusive regime. 
At a lower energy scale $E_2 \sim \frac{ D}{\xi^2}$, the density of 
states dips linearly to zero. (Here $D$ is the ``bare" spin diffusion constant, and 
$\xi$ is the quasiparticle localization length). This second energy scale 
marks the crossover 
from the diffusive to the localized regime. The ratio of the two crossover 
scales $\frac{E_1}{E_2}$ is exponentially large in the bare dimensionless
spin conductance, and can be quite large.

Note that our conclusion of the vanishing density of states is only 
superficially similar to the results of Nersesyan et. al.\cite{Tsvelik}. In 
particular, we argue that the localization length approaches a 
finite constant
as $E \rightarrow 0$. (In contrast, Nersesyan et. al. 
find a diverging localization 
length as $E \rightarrow 0$).

For a superconductor with spin rotation invariance 
but no time reversal (${\cal T}$),
there is again a logarithmic suppression of the density of states
in the diffusive regime. This accounts entirely for the leading logarithmic 
suppression of the spin conductance found earlier, 
and provides an explanation of it. In this case, we argue that 
in the localized phase, the density of states vanishes as $E^2$.

We provide an explicit verification of some of our general
results by exact non-perturbative calculations in one dimension
using supersymmetry techniques. 

\section{Diffusive regime} 
Consider first a two dimensional $d_{x^2 - y^2}$ superconductor with both spin rotation and 
${\cal T}$ invariance. In the absence of quasiparticle 
interactions, and
at scales larger than the elastic mean free path, 
the quasiparticle diffusion\cite{note0} is described by the replica field theory\cite{short}
\begin{equation}
\label{chiral}
S_{{\rm NL}\sigma{\rm M}} = \int d^2x \frac{1}{2g}{\rm Tr}\,(\nabla U \cdot
\nabla U^{\dagger}) -\eta {\rm Tr}\,(U + U^{\dagger})
\end{equation}
Here $U(x)$ are $2n \times 2n$ unitary matrices  $\in Sp(2n)$, and $\eta$ is a positive infinetisimal
that is introduced to enable calculation of the appropriate Green's functions.
The coupling constant $g$ is related to the 
spin conductance $\sigma_s$ by $\frac{1}{g} = \frac{\pi}{2}\sigma_s$.

The quasiparticle density of 
states at the Fermi energy is exactly proportional to the uniform spin susceptibility
which is the ``order parameter"\cite{short} for this field theory. Thus it is 
 given by
\begin{equation}
\rho = \lim_{n \rightarrow 0}\frac{\rho_0}{4n}\langle 
Tr\left(U^{\dagger} + U\right)\rangle
\end{equation}
where $\rho_0$ is the bare d.o.s ({\it i.e}, it's value
on the scale of the mean free path).
The limit $n \rightarrow 0$ is implied. 
As this is the ``order parameter" for the field
theory, quantum corrections to it 
can be obtained from the known results\cite{Zinn-Justin}
for the ``field renormalization" to one loop order.  
To perturbatively calculate
corrections to $\rho$, we write $U = 1 + i\phi - \frac{\phi^2}{2} + \cdots$
with $\phi$ belonging to the Lie algebra of $Sp(2n)$, and expand in 
powers of $\phi$. To quadratic order, the action is
\begin{equation}
S_0 = \frac{1}{2g}\int d^2x Tr (\nabla \phi)^2
\end{equation}
We may choose a basis ${T^a}$ for the Lie algebra to write
$\phi = \phi_a T^a$, and normalize the basis by some choice
$Tr(T^aT^b) = C\delta^{ab}$ with $C$ a positive constant. The action
then becomes
\begin{equation}
S_0 = \frac{C}{2g}\int d^2x \nabla \phi_a \cdot \nabla \phi_a
\end{equation}
(Summation over the index $a$ is implied in the above equations).
The matrices $T^a$ are traceless, and the susceptibility may be 
expressed to leading order as
\begin{eqnarray*}
\frac{\rho}{\rho_0}& = &\frac{1}{4n}\langle Tr(2-\phi^2) \rangle \\
& = &\frac{1}{4n}\langle 4n - Tr(T^a T^b)\phi_a \phi_b \rangle \\
& = & 1 - \frac{CN}{4n}G(0)
\end{eqnarray*}
In the last line, we have used the function $G(x)$ defined by
$\langle \phi_a(x) \phi_b(0) \rangle = G(x)\delta_{ab}$; $N$
is the number of linearly independent matrices $T^a$. For $Sp(2n)$,
$N= n(2n+1)$. It is understood that the $n = 0$ limit is taken at 
the end of the calculation. Considering now a finite system size,
we get
\begin{displaymath}
G(0) = \frac{g}{C}\int_{|k| > L^{-1}} \frac{d^2 k}{(2\pi)^2} \frac{1}{k^2}
\end{displaymath}
This gives
\begin{equation}
\label{dosft}
\frac{\rho - \rho_0}{\rho_0} = -\frac{gN}{8\pi n}\ln\left(\frac{L}{l_e}\right)
\end{equation}
Note that the constant $C$ has dropped out of this result (as it should). The
quantity $l_e$ is the elastic mean free path.
 
The same result is obtained in the ${\cal T}$ broken but spin rotation invariant case. This is described by
the $\frac{Sp(2n)}{U(n)}$ field theory\cite{short} with the action:
\begin{equation}
S = \int d^2x \frac{1}{2g} Tr \left((\nabla Q)^2 - \eta Q\sigma_z\right)
\end{equation}
where $Q = U^{\dagger}\sigma_z U$ with $U \in Sp(2n)$.  
The density of states is again the order parameter of this field theory, and is given by
\begin{equation}
\rho = \lim_{n \rightarrow 0}\frac{\rho_0}{2n}\langle 
Tr\left(Q\sigma_z \right)\rangle
\end{equation}
Calculation similar to the one above gives the result Eqn. \ref{dosft}
but with $N = n(2n+1) - n^2$ being
the number of independent massless fields. In either case, in the replica limit,
we get
\begin{equation}
\label{dosWL}
\frac{\rho - \rho_0}{\rho_0} = -\frac{g}{8\pi }\ln\left(\frac{L}{l_e}\right)
\end{equation}
Thus to leading order, the suppression of the density of states 
is independent of whether or not ${\cal T}$ is present.

The leading logarithmic correction to the spin conductance 
in two dimensions was evaluated
in Ref.\cite{short}:
\begin{equation}
\label{condWL}
\sigma_s(L) = \sigma_s^0 - \frac{1}{2\pi^2}\ln{\frac{L}{l_e}}
\end{equation}
where $\sigma_s^0$ is the bare spin conductance, and $l_e$ is the elastic
mean free path. If ${\cal T}$ is broken, then the correction is reduced by
a factor of two\cite{short}.

The spin conductance satisfies the Einstein relation 
$\sigma_s = \frac{D \rho}{4}$ with $D$ being the spin diffusion constant. 
(The factor of four arises from the spin 
of $1/2$.) Eqns. (\ref{dosWL}) and (\ref{condWL}) together with the relation 
$\frac{1}{g} = \frac{\pi}{2}\sigma_s$  imply a
logarithmic suppression of the diffusion constant at order $\frac{1}{g}$
when ${\cal T}$ is present. Without ${\cal T}$ invariance, there is no
suppression of the diffusion constant to this order.  

It is possible to understand this result in terms of a semiclassical
picture\cite{BB,Zirn} involving interfering trajectories. 
To that end, consider, quite generally, a lattice Hamiltonian 
for the quasiparticles
in a singlet superconductor
\begin{equation}
\label{BCS_c}
{\cal H} = \sum_{i,j}\left[t_{ij}\sum_{\sigma}c^{\dagger}_{i\sigma}c_{j\sigma} +
\Delta_{ij}c^{\dagger}_{i\uparrow}c^{\dagger}_{j\downarrow}+ 
\Delta^{*}_{ij}c_{j\downarrow}c_{i\uparrow} \right]
\end{equation}
where $i,j$ refer to the sites of some lattice.
Hermiticity implies $t_{ij} = t^{*}_{ji}$, 
and spin rotation invariance requires
$\Delta_{ij} = \Delta_{ji}$.   
It is useful conceptually to use the alternate representation in terms of a new set of 
$d$-operators defined by: $d_{i\uparrow} = c_{i\uparrow}, 
d_{i\downarrow}=c^{\dagger}_{i\downarrow}$. The Hamiltonian~Eqn.\ref{BCS_c} then
takes the form
\begin{equation}
\label{BCS_d}
{\cal H} = \sum_{ij}d^{\dagger}_i \left(\begin{array}{cc}t_{ij} & \Delta_{ij} \\
                                                \Delta_{ij}^{*} & -t_{ij}^{*} 
                                                               \end{array} \right) d_j
                                     = \sum_{ij}d_{i}^{\dagger}H_{ij}d_j
\end{equation}
Writing $t_{ij} = a^{z}_{ij}+ib_{ij},~~\Delta_{ij} = a^{x}_{ij}-ia^{y}_{ij}$
with $\vec{a}_{ij} = \vec{a}_{ji}$, real symmetric and $b_{ij} = -b_{ji}$, real antisymmetric,
we get 
\begin{equation}
H_{ij} = ib_{ij} + \vec{a_{ij}}.\vec\sigma
\end{equation}
Note that $SU(2)$ invariance requires $\sigma_y H_{ij}\sigma_y = -H^{*}_{ij}$.
This implies that the amplitude 
$iG_{ij,\alpha \beta} = \langle i\alpha |e^{-iHt}|j\beta\rangle \theta(t)$ for a 
$d$-particle to go from point $j$, (pseudo)spin $\beta$ to point $i$, spin $\alpha$
satisfies the relations:
\begin{eqnarray}
\label{Gsymm_1}
G_{ij,\uparrow \uparrow}(t) &= -G^*_{ij,\downarrow \downarrow}(t) \\
\label{Gsymm_2}
G_{ij,\uparrow \downarrow}(t) &= G^*_{ij,\downarrow \uparrow}(t) 
\end{eqnarray}
The Fourier transform of this amplitude is
\begin{eqnarray*}
G_{ij,\alpha\beta}(\omega +i\eta) & = & \int dt e^{i(\omega + i\eta)t} G_{ij, \alpha\beta}(t) \\
& = &  \langle i\alpha |\frac{1}{\omega - H + i\eta}|j\beta\rangle                                  
\end{eqnarray*}
The density of states at the Fermi energy may be obtained from this in the usual manner.
\begin{equation}
\rho = -\frac{1}{\pi}Im\left(\overline{G}_{ii,\uparrow\uparrow}(i\eta) + (\uparrow 
\leftrightarrow \downarrow \right)
\end{equation}

Consider now the return amplitude $G_{ii,\uparrow \uparrow}(t)$.
This can be written as a sum over all possible paths for this event. 
Consider in particular the contribution from the special class of paths
where the particle traverses some orbit and returns to the point $i$ in time $t/2$
with spin down, and then traverses the same orbit again in the remaining time and returns 
with spin up. This contribution to $iG_{ii,\uparrow \uparrow}(t)$ can be written
\begin{displaymath}
iG_{ii,\uparrow \downarrow}\left(\frac{t}{2}\right)iG_{ii,\downarrow \uparrow}\left(\frac{t}{2}\right)~~
= ~~-|G_{ii,\uparrow \downarrow}\left(\frac{t}{2}\right)|^2
\end{displaymath}
using the symmetry relation Eqn.\ref{Gsymm_2}. Now 
$|G_{ii,\uparrow \downarrow}(\frac{t}{2})|^2$ is just the probability 
for the event $i\uparrow \rightarrow i\downarrow$ in time $t/2$. For large $t$,
this is half the total return probability which $\sim \frac{1}{t}$ in two
dimensions
if the particles are diffusing. This leads to a logarithmic divergence in the 
density of states which may be cutoff by a finite system size. To be precise,
this gives
\begin{equation}
\frac{\delta \rho}{\rho_0} = -\frac{1}{\pi^2\rho_0 D}\ln\left(\frac{L}{l_e}\right)
\end{equation}
in agreement with the field theoretic result obtained earlier.

In addition, even to leading order 
the spin conductance for ${\cal T}$ invariant systems is suppressed 
further by the usual constructive interference between paths and their time reverse
which explains the larger suppression in that case.

\section{Localized regime}
Having established the presence of a singular suppression of the density of states
in perturbation theory, we now consider the opposite limit of strong disorder
when the system is localized. We  
show that the density of states vanishes at zero energy. To see this
heuristically, consider the Hamiltonian (\ref{BCS_c}) in the limit of strong on-site randomness and weak
hopping between sites. In the limit of zero hopping, the sites are all decoupled. 
At each site, the Hamiltonian in terms of the $d$-particles satisfies
the $SU(2)$ invariance requirement $\sigma_y H \sigma_y = -H^*$. This takes the form
$H = \vec a.\vec \sigma$ with $\vec a$ random. With ${\cal T}$ symmetry, we further
have $H = H^*$ implying $a_y = 0$. Considering now the case where the probability distribution
of $\vec a$ has finite, non-zero weight at zero, we see immediately that the disorder
averaged density of states vanishes as $E^2$ without ${\cal T}$ and as $|E|$
with ${\cal T}$. Now consider weak non-zero hopping. In the localized phase,
perturbation theory in the hopping strength should converge, and we expect to recover the 
single site results at asymptotically low energies. 

A more formal field theoretic version of this argument with the same conclusions is 
as follows. As we are concerned with the properties of the localized phase, we 
prefer to phrase the argument in terms of a supersymmetric field theory rather than the
replica version used before. In the localized phase, we expect that 
a strong coupling expansion of this field theory converges. This may be performed, as usual, by 
regularizing the sigma model on a lattice. The leading term in the strong coupling expansion
is the zero dimensional limit of the sigma model which is equivalent to the random matrix 
theory of Hamiltonians with these symmetries. In the random matrix limit, it is known\cite{Zirn}
that the density of states vanishes in the manner disussed above.

These results on the localized phase can be verified in great detail in one spatial
dimension for systems without ${\cal T}$. Consider a lattice Hamiltonian 
for the $d$-particles in one 
dimension. In the absence of disorder, we
take this to be of the form
\begin{equation}
{\cal H} = \sum_i -t(d_i^{\dagger} \sigma_z d_{i+1} + h.c) - \mu d_i^{\dagger}\sigma_zd_i 
\end{equation}

Now consider adding random terms to this Hamiltonian consistent with the 
required symmetry.
For weak disorder, we may just keep the modes near the two Fermi points
of the pure system. Linearizing the dispersion near these Fermi points, we arrive,
as usual, at a one dimensional Dirac theory with various sorts of randomness.
The resulting Hamiltonian can be written down on symmetry grounds as
\begin{equation}
H = -i\tau_z \partial_x + (\vec \eta_1(x).\vec \sigma)\left(\frac{1+\tau_z}{2}\right)
-(\vec \eta_2(x).\vec \sigma)\left(\frac{1-\tau_z}{2}\right) + t_0(x)\tau_y + (\vec t(x).\vec \sigma)\tau_x
\end{equation}
This is the most general Hamiltonian consistent with the symmetry $\sigma_y H \sigma_y = -H^*$
required by spin rotation invariance. (We have set the Fermi velocity to one). 
The $\vec \tau$ are Pauli matrices in the right 
mover/left mover space and $\eta_1,\eta_2, t_0, \vec t$ are random, independently distributed 
real variables. Green's functions of this Hamiltonian are generated by the action
\begin{equation}
S = \int dx \left[\overline{\psi}(iH + \omega)\psi + \xi^*(iH + \omega)\xi\right]
\end{equation}
where $\psi,\overline{\psi}$ are Grassmann variables, and $\xi$ is a complex scalar field. For a system
of finite size $L$, we impose periodic boundary conditions on 
all fields. The partition function $Z$ corresponding to this action is exactly equal to one for any
$L$ as the fermionic and bosonic integrals cancel each other.
Here $\omega$ is chosen to have a positive real part to ensure convergence of the bosonic 
integral.

The density of states can be obtained from the 
Green's function through
\begin{equation}
\rho(E) = -\frac{1}{\pi} Im~Tr \overline{G(E+i\eta)}
\end{equation}
where the overline indicates disorder averaging and 
\begin{equation}
G_{ab}^{\alpha\beta}(x,x';E +i\eta) = \langle a \alpha x|\frac{1}{E - H + i\eta}|b \beta x' \rangle
\end{equation}
It's disorder average can be expressed in terms of correlators of the either the Bose or Fermi 
variables:
\begin{equation}
\label{GF}
i\overline{G_{ab}^{\alpha\beta}(x,x';E +i\eta)} = \langle \psi_{a\alpha}(x)\overline{\psi}_{b\beta}(x')\rangle
=  \langle \xi_{a\alpha}(x)\xi_{b\beta}^*(x')\rangle
\end{equation}
(We have set $i\omega = E+i\eta$ in evaluating the correlators).
As we need 
the Green's function when $x = x'$, there is some subtlety on the relative ordering of $x$
and $x'$. The correct procedure\cite{BFZ} is to take a symmetrized form:
\begin{equation}
\label{dos1}
2\pi\rho(E) = Re\left[\langle \xi_{a \alpha}(x+\epsilon)\xi^*_{a\alpha}(x) + 
\xi_{a\alpha}(x-\epsilon)\xi^*_{a\alpha}(x) \rangle \right]
\end{equation}
where $\epsilon = 0^+$, and summation over $a, \alpha$ is implied. Precisely
the same expression with $\xi \rightarrow \psi$ holds in terms of the fermionic
variables as well.
The other physical quantity we will be interested in is the diffusion
propagator. This is defined , as usual, in terms of the Green's function by
\begin{equation}
P(x,x') = \sum_{ab,\alpha\beta}\overline{|G_{a\alpha,b\beta}(x,x';i\eta)|^2}
\end{equation}
Now the symmetry $\sigma_y H\sigma_y = -H^{*}$ can be used to show that
\begin{equation}
G_{ab}^{*\alpha\beta}(x,x';i\eta) = - (-1)^{\alpha+\beta}G_{ab}^{\bar{\alpha}\bar{\beta}}(x,x'; i\eta)
\end{equation}
where $\bar{\alpha} = 2$ if $\alpha = 1$ and vice versa.
Thus $P(x,x')$ may be written
\begin{eqnarray}
\label{Diffuson}
P(x,x')& = & -\sum_{ab,\alpha\beta} (-1)^{\alpha+\beta}\overline{G_{ab}^{\alpha\beta}
G_{ab}^{\bar{\alpha}\bar{\beta}}} \\
& = & \sum_{ab,\alpha\beta}(-1)^{\alpha+\beta}\langle\psi_{a\alpha}(x)\overline{\psi}_{b\beta}(x')
\xi_{a\bar{\alpha}}(x)\overline{\xi}_{b\bar{\beta}}(x')\rangle
\end{eqnarray}
We have chosen to write one Green's function
in terms of the fermions and one in terms of the bosons. This enables a calculation of the two
particle properties using the same formulation needed to calculate the one particle properties.

In the limit where $t^{\mu} = 0, \mu = 0,1,2,3$, the left/right moving
fields decouple for every realization of the disorder. Considering just one of them,
say the right-movers, we get the action
\begin{equation}
S = \int dx \left[\overline{\psi_1}\left(\partial_x + \vec \eta_1.\vec \sigma \right)\psi_1
+ \omega(\overline{\psi_1}\psi_1) +(\psi \leftrightarrow \xi)\right]
\end{equation}
We now average over the disorder assuming $\vec \eta_1$ to be distributed as
$P[\vec \eta_1] \propto exp\left[-\int dx \frac{(\vec \eta_1)^2}{2u}\right]$.
The resulting translationally invariant action can be interpreted as the coherent
state path integral of a zero dimensional quantum ``Hamiltonian" in terms of 
bose operators $b_{1\alpha} = (b_{1\uparrow}, b_{1\downarrow})$ and fermi
operators $f_{1\alpha} = (f_{1\uparrow}, f_{1\downarrow})$. Before doing that, we 
note that the fermionic fields actually satisfy
periodic boundary conditions. To get fermion fields that satisfy antiperiodic
boundary conditions, we may perform a change of variables
$\psi_1 \rightarrow \psi_1 e^{\frac{i\pi x}{L}}, 
\overline{\psi_1} \rightarrow \overline{\psi_1} e^{-\frac{i\pi x}{L}}$.
This adds a term $\int dx \frac{\pi}{L} \overline{\psi_1}{\psi}$ to the action.
Thus, we get 
\begin{eqnarray}
Z & = & STr e^{-Lh_R} \\
h_R & = & u\left(f_1^{\dagger}\vec \sigma f_1 + (f_1 \leftrightarrow b_1)\right)^2 +
\omega(f_1^{\dagger}f_1 + b_1^{\dagger}b_1)
\end{eqnarray}
(The subscript $R$ on $h$ is a reminder that this is for the right-moving fields alone). 
The supertrace operation $STr$ is defined through $STr {\cal O} = 
Tr \left((-1)^{f_1^{\dagger}f_1}{\cal O} \right)$. It is necessary to take the supertrace
to account for the extra term in the action coming from the change of the fermion
boundary conditions.
  
At zero $\omega$, it is clear that there is a triplet of states with zero energy:
the vacuum state with no particles which we denote $|0\rangle$, the state 
$f_{1\uparrow}^{\dagger}f_{1\downarrow}^{\dagger}|0\rangle \equiv |\uparrow \downarrow;0 \rangle$,
and the state $\frac{1}{\sqrt{2}}(b_{1\uparrow}^{\dagger}f_{1\downarrow}^{\dagger}
- b_{1\downarrow}f_{1\uparrow})|0\rangle \equiv |\uparrow;\downarrow \rangle$.
All other states have energies at least ${\cal{O}}(u)$. (Non-zero $\omega$ 
of course splits the energies of this zero energy triplet). 
Similar considerations apply to the left moving sector as well. Thus there are
a set of nine states all at zero energy at zero $\omega$ in the limit of
decoupled right/left sectors.

Now consider coupling the left/right moving sectors. The full action can also
be interpreted (after disorder averaging) as the coherent state 
path integral of a zero dimensional quantum Hamiltonian. In the limit where
the coupling is small, it is sufficient to project the interactions
induced between the two sectors to the nine dimensional space of the 
ground states of the two decoupled sectors. For simplicity, we assume that the $t_{\mu}$
are Gaussian distributed with
\begin{equation}
\overline{t_{\mu}(x)t_{\nu}(x')} = t^2\delta_{\mu \nu}\delta(x-x')
\end{equation}
To leading order then, the coupling between the two sites in the nine dimensional space
will be of order $\frac{t^2}{u}$. To derive the form of this coupling,
it is convenient to gauge away $\vec \eta_1$ and $\vec \eta_2$ by letting
$\psi_i \rightarrow U_i\psi_i, \overline{\psi_i} \rightarrow \overline{\psi_ i}U_i^{\dagger}$,
and similarly for $\xi$ with $U_i(x) = T_x\left[e^{i\int dx \vec \eta_i(x). \vec \sigma }\right]$
for $i = 1,2$. ($T_x$ is the $x$-ordering symbol.) We impose the condition
that $U_i(x = L) = 1$ to maintain the periodic boundary conditions. Note that the $U_i(x)$ are random
$SU(2)$ matrices. 
The full action can then be written
\begin{equation}
S = \int dx ~\overline{\psi} (\tau_z \partial_x + \omega)\psi + i (\overline{\psi}B(x)\tau^{+}\psi +
\overline{\psi}B^{\dagger}(x)\tau^{-}\psi) + 
(\psi \rightarrow \xi)
\end{equation}
Here $B(x) = U_1(x)(t_0 + i\vec t \cdot \vec \sigma)U_2^{\dagger}(x)$ is a random $2 \times 2$ matrix.
It is distributed according to
\begin{eqnarray*}
\overline{B^{\dagger}_{\alpha\beta}(x)B_{\gamma \delta}(x')}& = & t^2 e^{-u|x-x'|} \delta_{\alpha \delta}
\delta_{\beta \gamma} \\
\overline{B_{\alpha\beta}(x)B_{\gamma \delta}(x')}& = & t^2 e^{-u|x-x'|} (\sigma_y)_{\alpha \delta}
(\sigma_y)_{\beta \gamma}
\end{eqnarray*}
For large $u$, we may replace $ t^2 e^{-u|x-x'|} \rightarrow J \delta(x-x')$ with 
$J = \frac{t^2}{2u}$.
  It is now convenient to change variables
$\psi_{2\alpha} \rightarrow -\overline{\psi}_{2\alpha}$, $\overline{\psi}_{2\alpha}
\rightarrow \psi_{2\alpha}, \xi_{2\alpha} \leftrightarrow - \xi_{2\alpha}^*$. This 
changes the action to 
\begin{equation}
\int dx \left[\overline{\psi}\partial_x \psi + \xi^*\partial_x \xi
+ \left(-\overline{\psi_1} B(x) \overline{\psi_2}
- \psi_2 B^{\dagger}(x) \psi_1 - \xi_1^{*} B \xi_{2} + \xi_2 B^{\dagger} \xi_1 \right)
+ \omega (\overline{\psi}\psi + \xi^* \xi) \right]
\end{equation}
Under this change of variables, the expression Eqn. \ref{dos1} for the density of 
states remains unchanged (in the limit $\epsilon \rightarrow 0^+$). 

We may now perform the disorder average to get a translationally invariant action
which can be interpreted as the coherent state path-integral
of a zero dimensional quantum problem with bose operators $b_{a\alpha}$ and fermi
operators $f_{a\alpha} (a = 1,2; \alpha = \uparrow, \downarrow)$ and a ``Hamiltonian":
\begin{eqnarray}
\label{superH}
h & = & h_0 + h_{\omega} \\
h_0 & = & -J\left\{\left[(f_1^{\dagger}\sigma_yf_1^{\dagger})(f_2^{\dagger}\sigma_y f_2^{\dagger})
+ (f_1\sigma_y f_1)(f_2\sigma_y f_2) + 2(f_1^{\dagger}f_1 - 1)(f_2^{\dagger}f_2 - 1)\right] \right. 
\nonumber \\
&  & -2(b_1^{\dagger}b_1 + 1)(b_2^{\dagger}b_2 + 1) + \left[2(b_1^{\dagger}\sigma_y f_1^{\dagger})
(f_2^{\dagger}\sigma_y b_2^{\dagger}) + 2(b_1\sigma_y f_1)(f_2\sigma_y b_2) \right. \nonumber \\
&  & - 2(f_1^{\dagger}b_1)(f_2^{\dagger}b_2) - 2(b_1^{\dagger}f_1)(b_2^{\dagger}f_2))
\left. \right]\left.  \right\} \\
h_{\omega} & = & \omega(f^{\dagger}f + b^{\dagger}b)
\end{eqnarray}
Note that this Hamiltonian is non-hermitian. It's action on the nine dimensional subspace
is simplified by noting that $f_1^{\dagger}f_1 - f_2^{\dagger}f_2$, and $b_1^{\dagger}b_1
-b_2^{\dagger}b_2$ commute with $h$. Thus the six states 
$|\uparrow;\downarrow\rangle_1 \otimes |0\rangle_2, |0\rangle_1 \otimes |\uparrow;\downarrow\rangle_2,
|\uparrow \downarrow;0\rangle_1 \otimes |0\rangle_2, |0\rangle_1 \otimes |\uparrow \downarrow;0\rangle_2,
|\uparrow;\downarrow\rangle_1 \otimes |\uparrow \downarrow;0\rangle_2,
|\uparrow \downarrow;0\rangle_1 \otimes |\uparrow;\downarrow\rangle_2$ are immediately seen to be eigenstates
of the Hamiltonian. The first four have eigenvalues $4J + 2\omega$ and the last two have $4J + 4\omega$.
The action of $h$ on the remaining three states $|0\rangle_1 \otimes |0\rangle_2,
|\uparrow;\downarrow\rangle_1 \otimes |\uparrow;\downarrow\rangle_2,
|\uparrow \downarrow;0\rangle_1 \otimes |\uparrow \downarrow;0\rangle_2$ can be 
represented in terms of a $3 \times 3$ non-hermitian matrix:
\begin{equation}
\label{3matrix}
4 \left(\begin{array}{ccc}0 & -J & -J \\
                        J & 2J + \omega & J \\
                        -J & -J & \omega \end{array} \right)
\end{equation}
There is one eigenvalue $0$ and two eigenvalues $4J + 4\omega$.  With these
eigenvalues, it is easy to see that $Z = STr e^{-Lh} = 1$ for any system size $L$, as
required. We will also need 
to know the zero energy wavefunction. However due to the non-hermiticity
of the Hamiltonian, the left eigenvector($\vec V_L$) and the right eigenvector ($\vec V_R$)
are different. 
They are easily seen to be
\begin{eqnarray*}
\vec V_R & = & \frac{1}{1+\frac{\omega}{J}}\left[1+\frac{\omega}{J},  - 1,  1 \right] \\
\vec V_L & = & \frac{1}{1+\frac{\omega}{J}}\left[1+\frac{\omega}{J},  ~ 1,  1 \right] 
\end{eqnarray*}
We have normalized these so that $\vec V_L \cdot \vec V_R = 1$. There are some
subtle questions regarding the resolution of the identity in the basis of right (left)
eigenvectors of the Hamiltonian which are addressed at length in the Appendix.

\begin{figure}
\epsfxsize=4.0in
\centerline{\epsffile{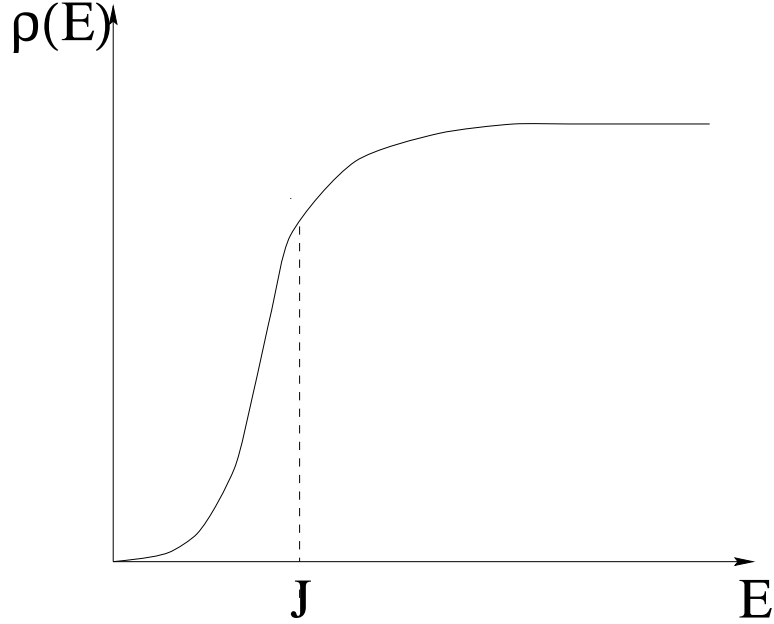}}
\vspace{0.05in}
\caption{Density of states in one dimension in the absence of time reversal symmetry;
the energy scale $J \sim \frac{1}{\xi}$ 
}
\vspace{0.04in}  
\label{dos1d}
\end{figure}

The expression Eqn. \ref{dos1} for the density of states can clearly be interpreted as the 
following expectation value: 
\begin{equation}
\label{dos2}
2\pi\rho  =  2Re \langle 2 + b^{\dagger}b \rangle
\end{equation}
where we calculate expectation values setting $i\omega = E + i\eta$. In the thermodynamic
limit only the zero energy state contributes, and the result is 
\begin{equation}
\label{dos3}
\frac{\pi}{2} \rho(E) = 1 - \frac{1-\frac{E^2}{J^2}}{(1 + \frac{E^2}{J^2})^2}
\end{equation}
Note that this vanishes as $E^2$ at small $E$, entirely consistent with
the general arguments presented earlier. For large $E$, this saturates at 
$\frac{2}{\pi}$ which is the ballistic result (see Figure \ref{dos1d}).

The diffusion propagator can also be calculated explicitly in this one dimensional
case. 
The calculation proceeds straightforwardly from Eqn.~\ref{Diffuson}. We first perform the change 
of variables $\psi_{2\alpha} \rightarrow -\overline{\psi}_{2\alpha}$, $\overline{\psi}_{2\alpha}
\rightarrow \psi_{2\alpha}, \xi_{2\alpha} \leftrightarrow - \xi_{2\alpha}^*$, and then interpret
the resulting correlator as an expectation value of an operator in the equivalent quantum 
problem. We find
\begin{eqnarray*}
P(x,x') & = & \langle T_x(O_1(x)O_2(x'))\rangle \\
O_1 & = & (f_{1\uparrow}b_{1\downarrow} - f_{1\downarrow}b_{1\uparrow}) + (f^{\dagger}_{2\uparrow}
b^{\dagger}_{2\downarrow} - f^{\dagger}_{2\downarrow}b^{\dagger}_{2\uparrow}) \\
O_2 & = & (f^{\dagger}_{1\uparrow}b^{\dagger}_{1\downarrow} - f^{\dagger}_{1\downarrow}b^{\dagger}_{1\uparrow})
- (f_{2\uparrow}b_{2\downarrow} - f_{2\downarrow}b_{2\uparrow}) 
\end{eqnarray*}
The expectation value is to be taken in the zero energy state. Consider $x> x'$ for definiteness.
Thus, we write
\begin{equation}
P(x,x') = \langle O_1 e^{-h(x-x')} O_2 \rangle
\end{equation}
We may evaluate this by inserting a complete set of states. As $O_2$ acting on the ground state is a state 
with energy $4J$ (when $\omega \rightarrow 0$), $P(x,x')$ decays as $e^{-4J(x-x')}$. The precise  
result is easily seen to be 
\begin{equation}
P(x,x') = 8 e^{-4J|x-x'|}
\end{equation}
Thus the localization length of the system is $\xi = \frac{1}{2J}$. In momentum space,
this becomes
\begin{equation}
P(q) = \frac{32J}{(4J)^2 + q^2}
\end{equation}
Note the difference in structure from conventional localization with a finite density 
of states where $P(q=0,\omega)$ has a pole at $\omega = 0$. In this problem, the density of
states vanishes and there is no pole.

The detailed calculation above of the one dimensional problem is strong evidence in 
support of our general assertions regarding the vanishing of the density of states in
the localized phase. In this case, the crossover from the constant to the vanishing density
of states occurs at an energy scale $J \sim \frac{1}{\xi}$ which is the energy scale for
the crossover from the ballistic to the localized regime. We expect that Eqn.~\ref{dos3}
is a universal scaling function for the density of states associated with this crossover.
In two dimensions (or in quasi one dimensional situations such as that 
considered in Ref. \cite{Carlos}), the crossover occurs 
between the diffusive and localized regimes at a scale $\frac{D}{\xi^2}$ (where $D$
is the diffusion constant). Again, this crossover is expected to be represented by a
universal scaling function for the density of states.

\section{Above two dimensions}
 We now turn to the situation above two dimensions where there
is the possibility of a spin metal to spin insulator transition.  The density of states 
is finite in the spin metal phase and vanishes on approaching the transition. Thus, in
contrast to usual Anderson localization, the density of states behaves as a 
conventional order parameter in these universality classes. The order parameter
exponent $\beta$ may be calculated within the $2+ \epsilon$ expansion. We find, to leading
order in $\epsilon$,
$\beta = \frac{1}{2}$ if ${\cal T}$ is present, and $\beta = 1$ without ${\cal T}$.
Right at the transition, the density of states vanishes with energy as
$\rho(E) \sim E^{\frac{1}{\delta}}$. The exponent $\delta = \frac{4}{\epsilon}, \frac{2}{\epsilon}$
with and without ${\cal T}$ respectively. 

\section{Discussion}
  In this paper, we have studied the behaviour of the quasiparticle density of states 
in a dirty $d_{x^2 - y^2}$ superconductor ignoring the quasiparticle interactions.
We showed the existence of a singular logarithmic suppression of the density of states
in the diffusive regime in two dimensions due to quantum interference effects. 
We then argued that in any dimension in the localized phase the density of states 
vanishes as $|E|$ if both spin rotation and ${\cal T}$ symmetry are present, and as
$E^2$ if spin rotation is the only symmetry. This was verified by a simple explicit
calculation in the latter case in one dimension using supersymmetry techniques.
Above two dimensions, we showed that the density of states is finite in the spin metal phase,
but vanishes on approaching the transition to the insulator. The corresponding
critical exponent was calculated in a $2+\epsilon$ expansion. These results are summarized in 
Table \ref{SC_loc}.

These results imply that the spin susceptibility, linear temperature coefficient
of the specific heat, and the tunneling density of states all have a logarithmic 
suppression as a function of temperature in the diffusive regime in two dimensions due to 
quantum interference effects. As pointed out in Ref.\cite{short}, inclusion of a 
Zeeman magnetic coupling drives the system into the usual unitary universality
class where there are no singular corrections to the density of states. Thus this logarithmic correction is 
killed by an external Zeeman field
(though not by a purely orbital magnetic field). Experimental verification of this effect
may be clouded somewhat due to the presence of quasiparticle interactions.  
We have shown elsewhere\cite{long} that in the diffusive regime, interaction
effects lead to a logarithmic Altshuler-Aronov suppression of the tunneling density of states
in the diffusive regime in two dimensions. This therefore adds to the quantum interference 
correction discussed
in this paper. In contrast, the specific heat and spin susceptibility are expected to get
logarithmic enhancements due to interactions in two dimensions in the diffusive regime\cite{MIT},
which too is killed by a Zeeman field.
They would thus compete with the quantum interference corrections. Nevertheless, 
if the interactions are weak, we expect that the quantum interference effects would dominate
leading to a logarithmic suppression of the spin susceptibility and specific heat, which
can be probed by applying an external Zeeman magnetic field.

The effect of interactions in the localized phase is a more delicate matter. Qualitatively,
repulsive interactions tend to favor the formation of local moments leading possibly to 
a divergent spin susceptibility and linear specific heat coefficient. This effect
will however compete with the vanishing density of states we have discussed above (which
tends to produce a vanishing spin susceptibility, etc). The ultimate fate of the 
localized phase in the presence of these two competing physical effects is a formidable 
problem that we will not attempt to answer here.

We are particularly grateful to Martin Zirnbauer for a most useful communication.  
We also thank Leon Balents, Ilya Gruzberg and Andreas Ludwig  for useful discussions. 
This research was supported by NSF Grants DMR-97-04005,
DMR95-28578
and PHY94-07194.

\begin{table}
\vspace{0.4in}
\caption{ Properties of the two different symmetry classes of superconductors with spin
$SU(2)$ symmetry. WL stands for weak localization.  
$\rho_{loc}(E)$ is the density of states in the localized phase. The last
column gives the critical properties of the density of states above two dimensions
as calculated in a $2+\epsilon$ expansion. The distance from the critical point is $(\delta g)$,
and $\rho_{cr}(E)$ is the density of states at the critical point.} 
\vspace{0.4in}
\begin{tabular}{|l|c|c|l|}    \hline
SYMMETRY          & WL  in $d=2$ & $\rho_{loc}(E)$ & CRITICAL PROPERTIES IN $d =2+\epsilon$ \\ \hline

Spin $SU(2)$ and ${\cal T}$       &$\frac{\delta \sigma_s}{\sigma_s} = 
-\frac{1}{2\pi^2\sigma_s}\ln\left(\frac{L}{l_e}\right)$& $|E|$ 
& $\rho(E = 0) \sim (\delta g)^{\frac{1}{2}}$ \\ 
 &$\frac{\delta \rho}{\rho} = -\frac{1}{4\pi^2\sigma_s}\ln\left(\frac{L}{l_e}\right)$ & 
&$\rho_{cr}(E) \sim |E|^{\frac{\epsilon}{4}}$ \\ \hline 

Spin $SU(2)$ and no ${\cal T}$  &$\frac{\delta \sigma_s}{\sigma_s} = 
-\frac{1}{4\pi^2\sigma_s}\ln\left(\frac{L}{l_e}\right)$& $E^2$ 
& $\rho(E = 0) \sim (\delta g)$ \\ 
 &$\frac{\delta \rho}{\rho} = -\frac{1}{4\pi^2\sigma_s}\ln\left(\frac{L}{l_e}\right)$ & 
&$\rho_{cr}(E) \sim |E|^{\frac{\epsilon}{2}}$ \\ \hline 
\end{tabular}
             
\label{SC_loc}

\end{table}

\appendix
\section{Resolution of the identity}
In this appendix, we discuss some subtle questions regarding the resolution of the identity
in the eigenbasis of the (super)Hamiltonian Eqn.~\ref{superH}. As all the subtleties
are associated with the three dimensional subspace spanned by $|0\rangle_1 \otimes |0\rangle_2,
|\uparrow;\downarrow\rangle_1 \otimes |\uparrow;\downarrow\rangle_2,
|\uparrow \downarrow;0\rangle_1 \otimes |\uparrow \downarrow;0\rangle_2$, we just focus on these 
three states. In this subspace, the Hamiltonian $h$ is represented by 
the $3 \times 3$ matrix Eqn.~\ref{3matrix}. The right eigenstates corresponding to the 
two eigenvalues $0$ and $4J + 4\omega$ are easily seen to be (in bra/ket notation)
\begin{equation}
|R_1 \rangle = a_1\left[\begin{array}{c}1+z \\ -1 \\ 1 \end{array}\right];~~
|R_2 \rangle = a_2\left[\begin{array}{c}0 \\-1 \\ 1 \end{array}\right]
\end{equation}
where $a_1, a_2$ are normalization constants, and $z = \frac{\omega}{J}$. ($|R_1\rangle$
has eigenvalue $0$, and $|R_2 \rangle$ has eigenvalue $4J + 4\omega$). Note that the
right eigenstates do not span the full three dimensional space. 
From the structure of the Hamiltonian Eqn.\ref{superH}, it is easy to see that from every 
right eigenstate $|R \rangle$,
a corresponding left eigenstate $\langle L|$ can be obtained by the operation
\begin{equation}
\label{leftrt}
\langle L| = ((-1)^{f_1^{\dagger}f_1}|R\rangle)^{T}
\end{equation}
where the symbol $T$ denotes taking the transpose.
The left 
eigenstates corresponding to $|R_1 \rangle$ and $|R_2 \rangle$ then are
\begin{equation}
\langle L_1 | = a_1\left[ 1+z, ~~1,~~1  \right];~~
\langle L_2 |=  a_2\left[ 0, ~~ 1, ~~1  \right]
\end{equation}
respectively, as can also be seen by direct calculation. The left 
eigenstates also do not span the full three dimensional space. 
Note that 
$\langle L_1| R_2 \rangle = \langle L_2 | R_1 \rangle  = \langle L_2 | R_2 \rangle = 0$.

To get a complete set of states, we need to supplement $\langle L_1|$ and 
$\langle L_2|$ by any other linearly independent bra vector $\langle L_3|$. 
It is convenient to choose this to be 
orthogonal to $|R_1\rangle$ and 
$(\langle L_2 | )^T$.
\begin{equation}
\langle L_3 | = a_3 \left[ -2, ~~ -(1+z), ~~ (1+z) \right]
\end{equation}
A corresponding right state $|R_3 \rangle$ can be defined using
Eqn. \ref{leftrt}:
\begin{equation}
|R_3 \rangle = a_3 \left[\begin{array}{c} -2 \\ 1+z \\ 1+ z \end{array}\right]
\end{equation}
Clearly the $|R_i \rangle (i = 1, 2,3)$ form a complete set of states 
(as do the $\langle L_i |$).
By construction, we have the relations $\langle L_3 |R_1 \rangle = \langle L_1 |R_3 \rangle
= 0$. We now impose the normalization conditions $\langle L_1 |R_1 \rangle = \langle L_2 |R_3 \rangle
= \langle L_3 |R_2 \rangle = \langle L_3 |R_3 \rangle = 1$. This fixes $a_1 = \frac{1}{1+z},
a_2 = -\frac{1}{1+z}, a_3 = \frac{1}{2}$. It is now possible to construct the resolution of
the identity
\begin{equation}
\label{unit}
1 = |R_1 \rangle \langle L_1 | + |R_2 \rangle \langle L_3 |+ |R_3 \rangle \langle L_2 |
- |R_2 \rangle \langle L_2 |
\end{equation}
This can be checked directly by it's action on any vector in the three dimensional 
space. 

This resolution of the identity can now be used to easily show explicitly 
that $Z = Str e^{-hL} = 1$ in the full nine dimensional space.
For the calculation of the density of states or the diffuson, 
we need to know the action
of the Hamiltonian $h$ on $|R_3 \rangle$. This is easily seen to be
\begin{equation}
h|R_3 \rangle = 4J(1+z) \left(|R_3 \rangle + |R_2 \rangle \right)
\end{equation}
Combined with the eigenvalue equation $h|R_2 \rangle = 4J(1+z) |R_2 \rangle$, 
this implies
that 
\begin{displaymath}
e^{-hL} |R_3 \rangle = e^{-4J(1+z)L}\left(|R_3\rangle - 4JLz(1+z)|R_2\rangle \right)
\end{displaymath}
In the limit $L \rightarrow \infty$, $e^{-hL}|R_3 \rangle \rightarrow 0$; similar 
considerations apply to $\langle L_3|$ as well. Calculation of any correlation function
is thus reduced, in the limit of infinite system size to a calculation in
the zero energy state with (right) eigenvector $|R_1 \rangle$.


\begin{references}
\bibitem{note0} As emphasized in Ref.\cite{short}, we mean here diffusion of the 
quasiparticle spin and energy densities.
\end{references}
\end{document}